\documentclass[11pt]{article}
\usepackage{graphicx}

\begin{document}

\title{Time and Reality of Worldtubes}
\author{Vesselin Petkov \\
Science College, Concordia University\\
1455 De Maisonneuve Boulevard West\\
Montreal, Quebec, Canada H3G 1M8\\
E-mail: vpetkov@alcor.concordia.ca}
\date{}
\maketitle

\begin{abstract}
In this essay, dedicated to the one hundredth anniversary of Hermann
Minkowski's talk ``Space and Time'', I argue that physicists should face the
issue of the reality of spacetime and worldtubes of physical objects for two
reasons. First, this issue is not a philosophical question, as some appear to
think, since the kinematical special relativistic effects would be impossible, as
will be demonstrated, if the physical objects involved in these effects did not
exist as four-dimensional worldtubes. Second, taking into account that
worldtubes are real four-dimensional objects provides an unexpected insight
into the origin of inertia, the nature of the force acting on a body supported in a
gravitational field, and possibly even the nature of quantum objects.
\end{abstract}

\vspace{.5cm}
\begin{flushright}
\textit{Things are never quite the way they seem}\linebreak Stannard Ridgway,
Lyrics ``Camouflage''
\end{flushright}

\section{Introduction}

Time is one of those profound features of the world which Nature has given us
as self-evident phenomena. Whenever we start thinking about the nature of time
we find ourselves in the situation perhaps best described by Saint Augustine
sixteen centuries ago: ``What then is time? Provided that no one asks me, I
know. If I want to explain it to an inquirer, I do not know'' \cite[p.
230]{augustine}.

Since ancient times time and existence have been closely linked -- it has always
been taken as self-evident that what exists, exists only at the present moment,
which means that what was in the future comes into existence by becoming
present and what existed as present goes out of existence by becoming past. In
other words, on this common-sense view, time objectively flows since each of its
moments consecutively comes into existence only once by instantaneously
becoming the moment `now'. This view, based solely on the fact that we realize
ourselves and the world only at the moment `now', which does not necessarily
imply that the world itself also exists only at this moment, is still widely accepted
even in the beginning of the 21st century.

For the first time this presentist view was questioned by the representatives of
the Eleatic school of philosophy twenty five centuries ago. Parmenides believed
that nothing can come into or go out of being (existence) because it would
contradict a basic postulate -- being exists, non-being does not exist -- that can
be deduced from what we perceive: ``there are signs aplenty that, being, it is
ungenerated and indestructible, whole, of one kind and unwavering, and
complete. Nor was it ever, nor will it be since now it is, altogether, one
continuous.'' \cite{barnes2001}. The Eleatics argued that their view of the world
as an eternal present followed from their basic postulate -- nothing can come
into being and become present, and nothing can go out of being, because
everything that exists exists as present and nothing else exists. If something
were to come into being it should come either from being or from non-being, but
neither of these alternatives is possible since being cannot become being (it is
already being), whereas non-being does not exist. By the same argument,
nothing can go out of existence. What is remarkable about the Eleatic
philosophy is that it presented the first example of a view about the world based
on more trust in reason than in our perceptions.

In fact, one does not need to agree with the Eleatic arguments to realize that not
only does the presentist view have difficulty explaining how something can come
into and go out of existence, but also that such a view appears to be
self-contradictory. It seems Aristotle was the first who came close to realizing
that self-contraction. He showed that the Eleatic philosopher Zeno arrived at his
paradox Dichotomy\footnote{An object moving from a point A to a distant point
B will never reach B since it will need an infinite amount of time first to travel
half of the distance AB, then half of the remaining half, and so on to infinity.},
designed to demonstrate the unreality of motion, because he explicitly
presupposed that space was divisible to infinity, but implicitly assumed that
time was not infinitely divisible. Aristotle wrote about Zeno's implicit
assumption: ``But this is false, because time is not composed of indivisible
nows, and neither is any other magnitude'' \cite[p. 161]{aristotle}. However,
when Aristotle discussed the nature of time itself he realized that he had to talk
about ``the present indivisible now'' \cite[p. 113]{aristotle} in order to avoid a
contraction in terms -- if the moment `now', which by definition is wholly
present, were divisible, it would contain past, present, and future parts. The very
fact that the founding father of logic, who single-handedly created the science of
logic, was led by the presentist view to the contradiction -- the present moment
is both divisible and indivisible -- implies that something is wrong with this
view.

Aristotle himself seems to have tried to identify the cause of that contradiction
and started to doubt whether the division of the moments of time into past,
present, and future reflected an objective fact or that division had something to
do with the mind: ``It might be wondered whether or not there would be time if
there were not mind'' \cite[p. 115]{aristotle}.

Saint Augustine also investigated the nature of time and like Aristotle faced the
same paradoxical situation about the duration of `now', but unlike him
concluded that the division of the moments of time into past, present, and future
does not reflect an objective feature of the world and therefore should belong to
the mind: ``What is by now evident and clear is that neither future not past
exists, and it is inexact language to speak of three times -- past, present, and
future [...] In the soul there are these three aspects of time, and I do not see
them anywhere else'' \cite[p. 235]{augustine}.

One might say that Aristotle and Saint Augustine had two options to avoid the
paradox with the duration of `now' -- either to assume that it is zero or that the
present moment has a finite duration, which is indivisible. On the presentist
view the first option is ruled out -- if the duration of `now' were zero it follows
that even the present moment would not exist, which means that no part of time
would exist. Most probably, Aristotle would have strongly objected to the second
option even because it would mean that Zeno would have succeeded in proving
that motion was impossible. If the moment `now' had a finite and indivisible
duration, this would challenge a basic assumption in the presentist view -- that
the physical objects and the world are three-dimensional -- an assumption that
has been regarded as self-evident since Aristotle \cite{3d}. To see the challenge
better, assume for the sake of the argument that `now' lasted ten seconds. This
would mean that physical objects and the world would not be three-dimensional
since they would exist \textit{at once at all moments} of the ten-second `now'
and would be therefore extended in time.

If the option of a finite, but indivisible `now' had been seriously examined before
the advent of special relativity, it would have been possible to analyze the nature
of time more rigorously and to distinguish two aspects of time whose mixture
has been causing a lot of confusion over the centuries -- the dynamical one,
according to which only one moment of time -- the constantly changing `now' --
exists, and the geometric (or static) one, which reveals the resemblance of time
to space (e.g. the resemblance of the interval between two moments of time to
the distance between two points in space). Such an analysis would have
strengthened the suspicion that the dynamical aspect of time is based
exclusively on the fact that we realize ourselves and the world at the constantly
changing present moment and uncritically assume that time objectively flows.
Then the question of whether the world itself also exists only at the moment
`now' could have been explicitly asked.

Another argument against the objectivity of the flow of time is that none of the
natural sciences deals with it. Moreover, special relativity and especially its
four-dimensional representation by Hermann Minkowski a hundred years ago
provided an explanation of this fact and of the resemblance of time to space --
space and time turned out to be different dimensions of an absolute underlying
reality -- spacetime. As all moments of time are entirely given as the forth
dimension of spacetime there is no objective flow of time. However, this does not
mean that there is no time. Simply, the dynamical aspect of time does not reflect
anything objective in the physical world. But time is very much real as the
temporal dimension of spacetime. The only explanation, compatible with
relativity, of our everyday experience, in which the feeling of time flow plays a
major role, appears to be Hermann Weyl's conjecture that it is the mind which
creates that feeling: ``The objective world simply \textit{is}, it does not
\textit{happen}. Only to the gaze of my consciousness, crawling upward along
the life line of my body, does a certain section of this world come to life as a
fleeting image in space which continuously changes in time'' \cite{weyl}.

In this essay I would like to draw the physicists' attention to the issue of the
reality of spacetime -- is spacetime just a mathematical four-dimensional space
or is it representing a real four-dimensional world? (By ``reality of spacetime'' I
mean a real four-dimensional world, not substantivalism.) In such a world
physical objects exist \textit{at once at all moments} of their histories in time as
four-dimensional worldtubes (spatially extended objects) or worldlines
(point-like particles). A hundred years after his profound insight we still owe
Minkowski a resolution of the issue of the reality of spacetime and worldtubes.
Many physicists do not see the need to address this issue despite that they freely
discuss extra dimensions and parallel universes. Some believe it is a
philosophical question. Others say this is not a real issue since relativity can be
equally represented in a three- and four-dimensional language. However,
neither of these reasons make the issue of the reality of spacetime disappear.
First, it is evident that the \textit{dimensionality} of the world and of physical
objects is not a philosophical question. Second, even if one agrees that the three-
and the four-dimensional representations of special relativity are equivalent, the
question of whether the world at the macroscopic scale is three- or
four-dimensional still needs an answer. Also, general relativity cannot be
adequately represented in a three-dimensional language.

Section 2 of the essay will provide the arguments supporting Minkowski's view
that the relativity ``postulate comes to mean that only the four-dimensional
world in space and time is given by phenomena'' \cite[p. 83]{minkowski}. It will
be demonstrated that the kinematical effects of special relativity are indeed
manifestations of the four-dimensionality of the world, which means that
physical objects are four-dimensional worldtubes in spacetime and that the flow
of time turns out to be minddependent as the Eleatics anticipated, Aristotle
suspected, and Saint Augustine and Weyl conjectured. The implications of the
reality of the worldtubes of physical objects for physics itself are explored in
Section 3 (for the origin of inertia), Section 4 (for the nature of the force acting
on a body supported in a gravitational field), and Section 5 (for the need of a
spacetime model of quantum objects).

\section{On the Reality of Spacetime and Worldtubes}

The very fact that the issue of the reality of spacetime has not been resolved so
far appears to suggest that the majority of physicists in the last hundred years
trusted more their senses than the arguments demonstrating that reality is not
the three-dimensional world of our perceptions. The resistance against
Minkowski's \textit{``postulate of the absolute world''} \cite[p. 83]{minkowski}
started immediately after his death. The depth of his idea of absolute
four-dimensional world does not seem to have been immediately realized as
evident from Sommerfeld's notes on Minkowski's paper: ``What will be the
epistemological attitude towards Minkowski's conception of the time-space
problem is another question, but, as it seems to me, a question which does not
essentially touch his physics'' \cite{sommerfeld}.

In the sixties Rietdijk \cite{rietdijk} and Putnam \cite{putnam} presented a
powerful argument demonstrating that relativity of simultaneity implies a
four-dimensional world. In fact, their argument was a special case of the more
general argument that appears to have led Minkowski to the idea of spacetime:
``We should then have in the world no longer space, but an infinite number of
spaces, analogously as there are in three-dimensional space an infinite number
of planes. Three-dimensional geometry becomes a chapter in four-dimensional
physics.'' \cite[pp. 79-80]{minkowski}. An infinite number of spaces means an
infinite number of sets of \textit{simultaneous} events corresponding to an
infinite number of observers in relative motion. It appears that it is precisely
relativity of simultaneity~--~different observers in relative motion have different
classes of simultaneous events and therefore different spaces -- that made
Minkowski realize that \textit{relative simultaneity and many spaces cannot
exist in a three-dimensional world}. And indeed, since a three-dimensional
world is defined in terms of \textit{absolute simultaneity} -- as one class of
simultaneous events -- relativity of simultaneity is \textit{impossible} in such a
world and does require a four-dimensional world with time as the fourth
dimension as Minkowski argued.

Strangely enough the relativity of simultaneity argument has not been taken too
seriously. Some physicists and philosophers do not accept the reality of the
four-dimensional world of relativity\footnote{In this context it makes no
difference whether the four-dimensional world is modelled by Minkowski
spacetime or another relativistic spacetime since any spacetime represents a
\textit{four-dimensional world} no matter flat or curved.}, but \textit{do not
explain} how relativity of simultaneity would be possible if the world were
three-dimensional.

An example is Stein's criticism of the Rietdijk-Putnam argument \cite{stein}. He
correctly pointed out that one could not talk about distant present events in
relativity but seemed to believe that he refuted their argument and some
philosophers agreed with him. What he refuted, however, is the presentist view
according to which what exists is a single class of (distant) present events -- a
single three-dimensional world defined as everything that exists
\textit{simultaneously} at the moment `now'. Stein criticized Rietdijk and
Putnam for arguing that relativity of simultaneity implies a four-dimensional
world, but explained neither how relativity of simultaneity would be possible if
the world were \textit{not} four-dimensional nor what the dimensionality of the
world according to relativity would be.

A similar objection against relativity of simultaneity as an argument for the
four-dimensionality of the world could be the fact that ``in special relativity, the
causal structure of space-time defines a notion of a `light cone' of an event, but
does not define a notion of simultaneity'' \cite{wald2006}. However, even if
such an objection were raised\footnote{Strictly speaking, such an objection
cannot be raised since it deals with causality, not existence.} it would, in fact,
directly rule out the three-dimensionalist view, which is defined in terms of
(absolute) simultaneity. More importantly, however, such an objection leads to
the four-dimensionality of the world even faster \cite{weingard},
\cite{petkov2007a}.

Despite that Rietdijk and Putnam did use the concept of distant present events
in their argument and that the causal structure of spacetime does not define a
notion of simultaneity, relativity of simultaneity remains a perfectly valid and
powerful argument in the context of the debate -- three-dimensional versus
four-dimensional world. The reason, I think, is obvious -- Minkowski, Rietdijk,
and Putnam started their analyses with the pre-relativistic three-dimensionalist
view in mind that space and the world constitute a single class of simultaneous
present events. Then it becomes immediately clear that if the world were
three-dimensional -- a single class of simultaneous distant present events --
simultaneity would be \textit{absolute} in contradiction with Einstein's special
relativity due to the fact that the notion of simultaneity is well-defined in a
three-dimensional world. It was then when Minkowski appears to have realized
that relativity of simultaneity is a manifestation of an absolute four-dimensional
world. Only in such a world one cannot talk about distant present events or,
more precisely, about simultaneous events.

Although the fact that no relativity of simultaneity is possible in a
three-dimensional world is sufficient, even taken alone, to prove the
four-dimensionality of the world, it is not the only argument. The real situation
is that the arguments for the reality of spacetime and worldtubes are
overwhelming. None of the kinematical special relativistic effects would be
possible if the physical objects involved in these effects were three-dimensional
\cite{petkov2005}--\cite{petkov2007a}. This demonstrates that the kinematical
relativistic effects are indeed manifestations of the four-dimensionality of the
world as Minkowski anticipated.

I will give just one example to demonstrate why, I think, the arguments, based
on specific relativistic effects, for the four-dimensionality of the world are
irrefutable~--~the relativistic length contraction of a meter stick would be
impossible if the meter stick were a three-dimensional object \cite{petkov2005},
\cite{petkov2007b}.

As a spatially \textit{extended} three-dimensional body is defined as ``all its
parts that exist \textit{simultaneously} at a given moment of time'' length
contraction is a specific manifestation of relativity of simultaneity. Then it
follows that while measuring the \textit{same} meter stick two observers in
relative motion measure \textit{two different three-dimensional meter sticks}
(two different sets of simultaneously existing parts of the meter stick) since the
two observers have different sets of simultaneous events. Therefore it is a
relativistic fact that the two observers measure two different three-dimensional
meter sticks, one of which is shorter. This is possible only if the worldtube of the
meter stick is a real four-dimensional object. Then the two three-dimensional
meter sticks, measured by the two observers in relative motion, are simply two
different three-dimensional ``cross-sections'' of it, which have different lengths.
What is the ``same meter stick'' is its worldtube.

Let me stress -- no length contraction\footnote{Length contraction was
experimentally tested, along with time dilation, by the muon experiment
\cite{muon}, \cite{ellis}.} would be possible if the meter stick's worldtube did
not exist as a four-dimensional object. If the meter stick were a
three-dimensional object, both observers would measure the \textit{same}
three-dimensional meter stick (the same set of simultaneously existing parts of
the meter stick), which would mean that the observers would have a common
(absolute) class of simultaneous events in contradiction with relativity.

\section{Worldtubes and Inertia}

We do not know why an accelerating body resists its acceleration. But by taking
into account that the worldtubes of physical bodies are real four-dimensional
objects the open questions of inertia and mass (as the measure of resistance a
body offers to its acceleration) can be viewed from an unexpected point of view.
As the worldtube of an accelerating body is deformed (is not geodesic) it seems
natural to assume that a four-dimensional stress arises in the deformed
worldtube of the body, which gives rise to a restoring force that resists the
deformation of the body's worldtube and tries to restore its geodesic shape. This
restoring force manifests itself as the inertial force. Calculations of the restoring
force (i) in terms of the four-dimensional stress tensor, (ii) in the case of the
classical electron, and (iii) by carrying out semiclassical calculations in quantum
field theory\footnote{The origin of the restoring force arising in a deformed
ordinary (three-dimensional) rod is explained in terms of electric forces acting
on the atoms of the rod, which are displaced from their equilibrium positions.
Similarly, the origin of the restoring force arising in the deformed worldtube of a
non-inertial object can be traced down to the elementary particles comprising
the object, which are also displaced from their equilibrium positions.}, all show
that the restoring force does have the form of the inertial force \cite[Ch. 10
]{petkov2005}. An inertial body whose worldtube is geodesic offers no
resistance to its motion because its worldtube is not deformed.

For centuries inertia has been one of the most profound puzzles in physics since
no one had suspected that its resolution might be linked to the dimensionality of
the world. I guess Minkowski would have been delighted to discuss the
possibility that inertia may turn out to be another manifestation of the
four-dimensionality of the world.

\section{Worldtubes and Gravitation}

General relativity provides a consistent no-force explanation of gravitational
interaction of bodies which follow geodesic paths, i.e. which are represented by
geodesic worldtubes. However, it is silent on the nature of the very force that has
been regarded as gravitational -- the force acting upon a body at rest in a
gravitational field.

Taking into account the reality of worldtubes makes the picture provided by
general relativity fully consistent. The worldtube of a body falling in a
gravitational field is geodesic and the body does not resist its fall since its
worldtube is not deformed. But when the body is prevented from falling its
worldtube is deviated from its geodesic shape.

The restoring force that arises in the body's deformed worldtube has the same
inertial origin as in the case of an accelerating body but in this case it manifests
itself as what is traditionally called the gravitational force. So, the force acting on
a body supported in a gravitational field is indeed \textit{inertial} \cite{rindler},
which naturally explains why ``there is no such thing as the force of gravity'' in
general relativity \cite{synge} and why inertial and gravitational forces (and
masses) are equivalent \cite[Ch. 10]{petkov2005}.

Another example of why physicists should take seriously the issue of the reality
of worldtubes, which can now display only staircase wit, is how naturally and
smoothly one arrives at general relativity when the implications of Minkowski's
idea of the absolute four-dimensional world are analyzed.

A conceptual analysis of Newton's gravitational theory could and should have
revealed, long before Einstein realized it, that a falling body offers no resistance
to its acceleration. This means that the body is not subjected to any gravitational
force, which would be necessary if the body resisted its fall. Therefore, the falling
body moves non-resistantly, by inertia. But how could that be since it
accelerates? Taking seriously the existence of worldtubes provides a radical
resolution -- the worldtube of the falling body should be curved (reflecting the
fact that the body accelerates), but not deformed (accounting for the fact that
the body does not resist its acceleration). Such a worldtube can exist only in a
curved spacetime where the geodesic worldtubes of bodies moving by inertia are
curved but not deformed.

\section{Wordlines and Quanta}

No one knows what the quantum object (e.g. an electron or photon) is. What is
worse is that the standard interpretation of quantum mechanics tells us that we
cannot say or even ask anything about the quantum object between
measurements. In this sense, I think, Einstein was right that quantum
mechanics is essentially incomplete. It is unrealistic to assume that an electron,
for example, does not exist between measurements. But if it exists, it is
something and we should know what that something is.

Although relativity does not fully apply at the quantum level since its equations
of motion manifestly fail to describe the behaviour of quantum objects,
spacetime does seem to be the arena of the quantum world as well. Then the
question is: ``Are quantum objects also worldlines in spacetime?'' The answer is
well known -- definitely not.

This can be demonstrated in the case of interference experiments performed
with single electrons \cite{tonomura}. In such double-slit experiments
accumulation of successive single electron hits on the screen builds up the
interference pattern that demonstrates the wave behaviour of \textit{single}
electrons. When looking at the screen, every single electron is detected as a
localized entity and the natural question is whether the electron was such an
entity before it hit the screen. Careful conceptual analysis indicates that it is not
quite clear what is meant by "such an entity". Our intuition leads us to assume
that if the electron hits the screen as a localized entity, it is such an entity at
\textit{every} moment of time, which means that the electron exists
\textit{continuously} in time as a localized entity. But if this were the case, every
single electron would behave as an ordinary particle and should go only through
one slit and no interference pattern would be observed on the screen. Therefore,
the electron cannot be a localized entity at all moments of time, i.e. it cannot be
a worldline in spacetime.

The paradox -- every single electron must go through both slits (in order to hit
the screen where the ``bright'' fringes of the interference pattern form) but is
always detected on the screen as a localized entity -- may turn out to be caused
by the same implicit assumption that caused Zeno's paradox Dichotomy. Zeno
assumed something about space (infinite divisibility) but did not assume the
same thing about time. It seems we have been making exactly the same mistake
when we arrive at the quantum paradoxes -- explicitly regarding an electron as
localized in space, but implicitly assuming that it is continuously existing in
time, i.e. not being localized in time.

Perhaps the best way to envisage an electron which does not exist continuously
in time is to imagine that its worldline disintegrates into its constituent
four-dimensional points. The Compton frequency implies that for one second an
electron will be represented by $10^{20}$ such points. When an electron is not
measured it is \textit{actually} everywhere in the spacetime region where its
wavefunction is different from zero, because its constituents are scattered all
over that region; so such an electron naturally goes through both slits. When the
first four-dimensional point of an electron falls in a detector it is trapped there
due to a jump of the boundary conditions and all its consecutive points also
appear in the detector, which means that an electron is always measured as a
localized entity.

In this spacetime model of the quantum object the probabilistic behaviour of
quantum objects and the relativistic forever given spacetime picture of the world
do not contradict at all -- the \textit{probabilistic distribution} of the
four-dimensional points of an electron in the spacetime region where the
electron wavefunction is different from zero is \textit{forever given} in
spacetime.

In these desperate times in quantum physics it is worth searching for a
spacetime model of quantum objects and testing the idea \cite{nasko} that they
do not exist continuously in time. Moreover, this idea provides a paradox-free
interpretation of quantum mechanics\footnote{For a more detailed conceptual
account of the idea that quantum objects may exist discontinuously in time see
\cite[Ch. 6]{petkov2005}.} and may even lead to its further development.

\section*{Conclusion}

A hundred years after Minkowski presented his paper ``Space and Time'' we
still owe him a definite answer to the question of the reality of spacetime and
worldtubes. In this essay I argued that a rigorous conceptual analysis of the
kinematical relativistic effects demonstrates that spacetime represents a real
four-dimensional world in which physical bodies are worldtubes. Taking into
account the reality of worldtubes sheds light on the origin of inertia, on the
nature of the force acting on a body supported in a gravitational field, and even
on what the quantum object might be.

In this essay I also wanted to emphasize again the crucial role of conceptual
analyses for (i) any advancement in fundamental physics and (ii) deep
\textit{understanding} of physical theories, because it appears that such
analyses are regarded by some physicists as old-fashioned and even belonging to
philosophy. The history of the fundamental breakthroughs in physics, however,
convincingly demonstrates that conceptual analyses are physics at its best.

\end{document}